# Metal-organic chemical vapor deposition of MgGeN$_2$ films on GaN and sapphire

Chenxi Hu[1], Vijay Gopal Thirupakuzi Vangipuram[2], Christopher Chae[3], Ilteris K. Turan[1], Nichole Hoven[4], Walter R.L. Lambrecht[1], Jinwoo Hwang[3], Yumi Ijiri[5], Hongping Zhao[2,3,*], Kathleen Kash[1,‡]

[1] Department of Physics, Case Western Reserve University, Cleveland, OH 44106, USA
[2] Department of Electrical and Computer Engineering, The Ohio State University, Columbus, OH 43210, USA
[3] Department of Materials Science and Engineering, The Ohio State University, Columbus, OH 43210, USA
[4] Swagelok Center for Surface Analysis of Materials, Case Western Reserve University, Cleveland, OH 44106, USA
[5] Department of Physics and Astronomy, Oberlin College, Oberlin, OH 44074, USA

[*] Email: Zhao.2592@osu.edu
[‡] Email: kathleen.kash@case.edu

**Abstract**

MgGeN$_2$ films were synthesized using metal-organic chemical vapor deposition on GaN/c-sapphire templates and c-plane sapphire substrates. Energy-dispersive X-ray spectroscopy was used to estimate the cation composition ratios. To mitigate magnesium evaporation, the films were grown at pyrometer temperature 745 °C with a wafer rotation speed of 1000 rpm. Growth rates were determined by fitting energy-dispersive X-ray spectroscopy spectra to film thicknesses using NIST DTSA-II software. The thickness estimates determined by this method were consistent with scanning transmission electron microscopy measurements done for selected samples. Scanning electron microscopy images revealed faceted surfaces indicative of a tendency toward three-dimensional growth. X-ray diffraction spectra confirmed that the films were highly crystalline and exhibited preferential orientation in alignment with the substrate. Atomic force microscopy measurements show that film thicknesses are consistent across samples grown on both GaN templates and sapphire substrates, with typical roughnesses around 10 nm. Transmittance spectra of films grown on double-side-polished sapphire substrates yielded band gaps of 4.28 ± 0.06 eV for samples exhibiting close-to-ideal stoichiometry. Comparison of the measured spectra with ab initio calculations are in good agreement both near the band gap and at higher energies where excitation is into higher-lying bands. These findings provide insight into the growth and characterization of $MgGeN_2$, contributing to the development of this material for potential applications in optoelectronics and power electronics.

## I.  Introduction

A large family of ternary heterovalent II-IV-$N_2$ compounds shares a close relationship with III-N compounds regarding their structural characteristics. Theoretically, one can conceptualize the ideal ground state of these II-IV-$N_2$ structures as an evolution from a wurtzite III-N lattice, involving a substitution process where every pair of group-III atoms in the crystal lattice is replaced by an ordered combination of one group-II element atom and one group-IV element atom such that every nitrogen atom is bonded to two of each cation species in order to satisfy the octet rule. This hypothetical process of atomic substitution provides insight into the fundamental structural similarities and differences between these two classes of compounds [1,2,3]. One notable difference is that for the II-IV-$N_2$ family the band gaps may be modified by introducing disorder on the cation sublattice without changing the composition [4].

Much recent interest in the Zn-IV-$N_2$ family, especially $ZnGeN_2$ and $ZnSnN_2$ [5,6,7], has been prompted by optoelectronics applications such as solar cells, especially for $ZnSnN_2$, and for extending the wavelength range for efficient nitride-based LEDs in combination with the III-nitrides [8,9]. $MgGeN_2$ and $MgSiN_2$ present interesting opportunities to design systems with ultrawide band gaps. These compounds and their alloys demonstrate considerable flexibility in achieving closer lattice matches with GaN substrates. In addition, growth on AlN substrates results in compressive stress in the films, which avoids the tendency toward cracking that tensile stress produces.

$MgGeN_2$ is a direct band gap material with a predicted gap of 4.11 eV, as shown in Figure 1 [10,11,12]. $MgSiN_2$ has a predicted direct band gap of 6.3 eV and an indirect gap of 5.8 eV [12]. Alloy systems combining $MgSiN_2$ with $MgGeN_2$, GaN, or AlN could span from approximately 200 to 300 nm wavelengths, covering much of the ultraviolet C (UV-C) range [10,12]. When grown on GaN substrates, $MgGeN_2$ experiences a lattice mismatch of roughly 4% in the a-direction and about −0.5% in the b-direction. This situation contrasts with the considerably smaller lattice mismatch of around -0.5% in the b-direction [13, 14].

Advancements in growth methodologies for $MgGeN_2$ and the exploration of its intrinsic properties are still in the early stages. $MgGeN_2$ powder was synthesized from the precursor $Mg_2Ge$ [15, 16]. To date, the growth of films of $MgGeN_2$ has been reported in only a few studies, using molecular beam epitaxy (MBE) [17] and radio frequency (RF) sputtering [18]. MBE studies of films grown on Si (111) and $Al_2O_3$ (0001) substrates have shown the formation of periodic growth mounds on the surfaces. On $Al_2O_3$ (0001) substrates, the stoichiometric $MgGeN_2$ thin films were grown using a nitrogen plasma source at 750°C, after high temperature annealing of the substrate in an ammonia atmosphere. On Si (111) substrates, a buffer layer of AlN was first deposited, followed by the growth of $MgGeN_2$. A formalism for dynamical roughening applied to these mounds predicted an average spacing of 235 nm. This effect has been attributed to diffusion bias:

an adatom diffusion current on a vicinal surface caused by an asymmetry in the kinetics of adatom attachment to steps [17].

We report here the synthesis of MgGeN$_2$ using metal-organic chemical vapor deposition (MOCVD). Films were grown on GaN/c-sapphire templates and c-plane sapphire substrates at a pyrometer temperature of 745 °C with a wafer rotation rate of 1000 rpm. The precursors used include Cp$_2$Mg and MeCp$_2$Mg, along with germane and ammonia, with N$_2$ used as the carrier gas. Stoichiometry, measured by energy-dispersive X-ray spectroscopy (EDS), was achieved with a Mg precursor-to-germane flow rate ratio between 7 and 13. Growth rates of approximately 100 nm/h using the precursor MeCp$_2$Mg, and approximately 30-40 nm/h using the precursor Cp$_2$Mg, were measured by fitting the EDS spectra to film thicknesses using NIST DTSA-II software [19, 20, 21]. These results are consistent with scanning transmission electron microscopy (STEM) EDS measurements, done on selected samples. Scanning electron microscopy (SEM) images of MgGeN$_2$ films show faceting. X-ray diffraction (XRD) spectra confirmed the films' high degree of crystallinity. Film thicknesses and stoichiometries were consistent across samples grown on GaN templates and sapphire substrates, with roughnesses around 10 nm as measured by atomic force microscopy (AFM).

## II. Experimental details

The MgGeN$_2$ films were grown using a custom dual-chamber MOCVD system built by Agnitron, Inc. Commercially available GaN/c-sapphire templates from Lattice Power, with threading dislocation densities of $4\times10^8$ cm$^{-2}$, were co-loaded with c-sapphire substrates. The AFM root-mean-square (RMS) roughness for both templates were under 1 nm.

As noted above, we used two different magnesium precursors: bis-cyclopentadienyl magnesium ($Cp_2Mg$), and bis(methylcyclopentadienyl) magnesium ($MeCp_2Mg$), along with ammonia and two concentrations (2% and 0.02%) of germane ($GeH_4$). Nitrogen was used as the carrier gas. The carrier nitrogen flow rate was 2000 sccm, and the flow rate of nitrogen through the bubbler was 450 sccm. The growth parameters, simulated thicknesses, and atomic compositions of various $MgGeN_2$ samples grown under different conditions are shown in Table 1. For the 2% germane source, the germane flow was pulsed in 1-minute cycles consisting of three phases: linear ramp-up and ramp-down rates (together constituting the t_ON period), and an off-time t_OFF. A schematic representation of the pulsed-mode growth sequence is shown in Figure 2. For the 0.02% germane source, the germane flow rate was kept constant.

The atomic percentages of Mg, Ge, and N in the $MgGeN_2$ films were measured by EDS with an Apreo SEM, which was also utilized for field emission scanning electron microscopy (FESEM) to investigate surface morphologies. Film thicknesses were determined by fitting the EDX data, using NIST DTSA-II software, for films grown on GaN templates. Here the determination of the thickness of the $MgGeN_2$ layer was based on the relative Ga intensity originating from the underlying GaN layer. For one sample the film thickness was also measured by cross-sectional SEM imaging. A Bruker Icon 3 AFM was used to measure the films' roughness. XRD measurements were performed with a Rigaku Ultima IV in standard powder diffraction with Bragg-Brentano geometry and with a Rigaku SmartLab in Parallel Beam geometry with a 0.8 mm collimator and equipped with a HyPix-3000 detector measuring in 1D mode. The acquisitions on the SmartLab scanned 20-100º 2Θ with a step size of 0.01º and scan speed of 2.5º/min. Both instruments used Cu Kα radiation. STEM was carried out using a Thermo Scientific probe-corrected Themis-Z STEM. Transmittance measurements were done using a Cary 5000 spectrophotometer.

### III. Results and discussion

**A. MgGeN$_2$ cation stoichiometry and growth rates of pulsed growth mode**

Table 1 lists the growth parameters, simulated thicknesses, and atomic compositions of the MgGeN$_2$ samples used in this study. The R$_{II/IV}$ ratio represents the Mg:Ge molar flow rate ratio. To achieve the targeted 1:1 atomic ratio of Mg and Ge, our studies indicate that the R$_{II/IV}$ ratio should be adjusted to between 7 and 13 under the growth conditions used here.

Figure 3 shows the relationship between the Mg/Ge EDS ratio and the R$_{II/IV}$ ratio along with the growth temperature. As expected, lower growth temperatures favor higher Mg incorporation at a given flow rate ratio, likely due at least in part to reduced Mg evaporation from the growing surface.

The atomic percentages of Mg, Ge, and N in the MgGeN$_2$ films and the film thicknesses, are shown in Table 1. The estimations of the film thicknesses were based on fitting the EDS spectra, including the Ga peak originating from the GaN layer beneath the MgGeN$_2$ layer for the samples grown on GaN templates. We typically performed SEM-EDS thickness measurements in a 2 × 2 μm region near the center of each sample. While this localized approach introduces some uncertainty—given the surface roughness and radial variation in film thickness—consistency in how and where the measurements are made allows for reliable relative comparisons across different samples. Hence, although the absolute thickness values may deviate if examined over larger or different areas, the simulated EDS spectra remain useful in providing an analytical strategy and in offering a comparative sense of thickness from one sample to another. For example, when comparing the Ga intensities of samples B and D in Table 1, it is clear that sample D has a slightly elevated Ga ratio relative to sample B, resulting in the estimate of a lower $MgGeN_2$ film thickness.

Figure 4 shows cross-sectional STEM EDS of Sample E, highlighting the distribution of Mg, Ge, and N within the film. For this sample, the atomic ratio for Mg to Ge is stoichiometric within measurement uncertainty. The decrease in nitrogen and the increase in cation concentrations is an artifact of the measurement. Figure 5 (a) shows both the simulated and measured EDS results, which are consistent with each other. Figure 5 (b) presents the measured and fitted EDS spectra for sample E and F. The spectra reveal the same nitrogen peak intensity for both samples, while the peak intensities for Ga, Mg, and Ge differ, yielding a different thickness for each of these films.

**Calculated and measured growth rates**

Detailed modeling of the growth rate is extremely complex for the III-nitrides [22]. It is considerably more complicated in the case of the II-IV-nitrides, with the incorporation of two cation precursors. However, even the results of a simple model that circumvents consideration of both the gas phase and surface reactions and kinetics can help give insight into and guidance for the growth process. We start with the following equation for the arrival rate per unit area of a precursor molecule to the growing surface [23]:

$$J = \frac{D_{AB} * P^*_{pr}}{R \delta_0 T} \quad (1)$$

Here $D_{AB}$ is the diffusion rate of species A in species B (or vice versa) for diffusion in a bimolecular gas. $P^*_{pr}$ is the partial pressure of the rate-limiting precursor molecule in the growth chamber, R is the gas constant, T is the temperature in K, and $\delta_0$ is the boundary layer thickness [23]:

$$\delta_0 = 4\left(\frac{v}{\omega}\right)^{1/2} \tag{2}$$

Here $\omega = 2\pi n$, where n is the rotation rate of the susceptor, and $v$ is the kinematic viscosity of $N_2$.

The partial pressure of the rate-limiting precursor molecule in the chamber is:

$$P_{pr}^* = P\frac{P_{pr}f_{source}}{P_{source}f_{total}} \tag{3}$$

Here P is the chamber pressure, $P_{pr}$ is the partial pressure of the precursor at its source, $P_{source}$ is the pressure in the precursor's source, $f_{source}$ is the flow rate from the precursor's source, and $f_{total}$ is the total flow rate into the reaction chamber. If the rate-limiting species is germane, $P_{pr}/P_{source}$ is the concentration of germane in the germane/hydrogen compressed gas cylinder. If the rate-limiting species is Mg, $P_{pr}$ depends on the bubbler temperature. The total flow rate is $f_{total} = f_{bubbler} + f_{NH3} + f_{N2} + f_{germane}$, where $f_{bubbler}$ is the flow rate of gas through the bubbler, $f_{N2}$ is the flow rate of the carrier gas, and $f_{germane}$ is the flow rate from the germane-hydrogen gas cylinder. The partial pressure of $Cp_2Mg$ versus the bubbler temperature is established experimentally [24]. The partial pressure of $MeCp_2Mg$ was calculated here using the Clausius-Clapeyron equation with the boiling point of 333 K at 1.0 torr and the vapor pressure of 0.25 torr at 296 K, as provided by the manufacturer Dockweiler.

We use the following equation for $D_{AB}$ [25]:

$$D_{AB} = \frac{3}{8}\frac{\left(\frac{\pi kT}{M_{AB}}\right)^{1/2}kT}{P\pi\sigma_{AB}^2}\frac{f_D}{\Omega_D} \tag{4}$$

Here $f_D = 1$ is a correction factor of the order of unity and is usually set to 1. The factor $\Omega_D$ is a dimensionless collision integral that depends on the choice of the interaction of the molecules (via, for example, the Lennard-Jones potential parameters), and is also of order 1. $M_{AB}$ is the combined molecular weight for mixtures or compounds [26]; $M_{AB} = 2[(1/M_A) + (1/M_B)]^{-1}$, where $M_A$ and $M_B$ are the molecular weights of components A and B, and $\sigma_{AB}$ is the average diameter of the interacting molecules; $\sigma_{AB} = (\sigma_A + \sigma_B)/2$. The molecular weights of Cp$_2$Mg (chemical formula C$_{10}$H$_{10}$Mg) and Me(Cp)$_2$Mg (chemical formula C$_{12}$H$_{14}$Mg) are 154.5 g/mol and 182.6 g/mol, respectively.

In estimating $\sigma_{AB}$ we consider the precursors' interactions with the N$_2$ carrier gas. We take the diameter $\sigma_A$ of a nitrogen molecule to be 0.38 nm [25] and neglect the uncertainty in this quantity. In estimating the average diameters of the precursor molecules, we note that both Mg precursors consist of two 5-carbon rings linked by the Mg atom. The molecular diameters for benzene and cyclohexane of 0.53 and 0.62 nm, respectively [26]. For the molecular diameter of Cp$_2$Mg we take the molecular diameter of benzene, calculate its molecular volume, multiply by two, add the estimated volume of a Mg atom, with an atomic radius of 0.28 nm, and assume a spherical molecule. We use the same procedure for MeCp$_2$Mg but use the molecular diameter of cyclohexane instead of benzene to account for the CH$_3$ attachment to each ring. The molecular diameter is $\sigma_B = 0.68 \pm 0.17$ nm for Cp$_2$Mg and $\sigma_B = 0.79 \pm 0.19$ nm for MeCp$_2$Mg. The molecular diameter of GeH$_4$ is $\sigma_B = 0.15 \pm 0.02$ nm.

Finally, the growth rate in m/s is J*A*V$_{cell}$/4, where A is Avogadro's number, V$_{cell}$ is the MgGeN$_2$ unit cell volume of 0.193 nm$^3$ [14], and the factor 4 arises from the consideration that there are four Mg atoms per Pna2$_1$ unit cell.

Figure 6 illustrates the relationship between the experimental growth rates and the growth rates calculated presuming that either the Mg precursor or germane is the rate-limiting species. For these calculations we have assumed that either the Ge or the Mg is the growth-rate-limiting species, that the delivery of the atom by the precursor to the growing surface is 100% efficient, and that there is no re-evaporation of the species from the surface or decomposition of the film during growth. The calculated growth rates based on the Mg precursors are approximately 5 to 6 times higher than the experimental growth rates, and for Ge, the calculated growth rates are 3.5 times higher. The discrepancies between the calculated and measured growth rates may also be partly due to the assumptions for the values of the molecular diameters. In any case, this simple model provides an important order-of-magnitude estimate of expected growth rates and offers predictions for trends as growth parameters are varied.

**B. MgGeN$_2$ surface morphologies and crystallinity**

In this section, we examine the morphology and crystallinity of the MgGeN$_2$ samples using SEM imaging, AFM, and XRD measurements.

We compare the surface morphologies of samples D, E, and F to understand the effect of different growth conditions. Figure 7 (a) presents the SEM image of sample D, grown at 800°C using MeCp$_2$Mg as the precursor. The growth rate for this sample is measured at 45 ± 1.5 nm/h, which is significantly lower than that of samples E and F, suggesting that high temperatures are less favorable for MgGeN$_2$ growth. In contrast, Figures 7 (b) and (c) show the surface morphologies of samples E and F, which were grown at a lower temperature of 720°C with the same R$_{II/IV}$ ratio of 7. The only difference between these two samples is the magnesium source bath temperature: 30°C for sample E and 40°C for sample F. This adjustment in the bath temperature resulted in a higher magnesium flow rate for sample F, accompanied by a proportional increase in the germane flow rate to maintain the same II/IV ratio. The SEM images of both samples reveal surfaces composed of aligned facets, but increased Mg flow rate in Sample F contributes a higher growth rate. For subsequent experiments, we adopted the higher magnesium bath temperature of 40°C as the standard.

Next, we focus on the samples grown using $Cp_2Mg$ as the precursor on both GaN and sapphire substrates. Figures 7 (d) and (e) display the surface morphologies of these samples, grown under identical parameters with an $R_{II/IV}$ ratio of 9. The key difference between the two samples is the growth time: sample J had a longer growth duration of 2 hours, while sample I was grown for a shorter time, 1 hour. Due to the lower vapor pressure of $Cp_2Mg$ compared to $MeCp_2Mg$, these samples exhibited a reduced growth rate. Sample I formed a $MgGeN_2$ layer with a thickness of 36±1 nm, while Sample J achieved a thickness of 68±2 nm. Unlike the aligned facets observed in samples E and F, the SEM images of samples I and J show no specific facet alignment but rather some 3D crystals scattered on a relatively flat surface.

Using EDS, as shown in Figures 8 (a) and (b), we determined that both the flat surface and the 3D crystal structures are composed of $MgGeN_2$. Additionally, Figures 7 (f) and (g) provide plan view SEM images of sample K, which has a slightly higher $R_{II/IV}$ ratio of 2.2% compared to sample J. The SEM images of sample K show a notably flatter surface, devoid of the larger 3D crystals observed in sample J. For sample K, both GaN and c-sapphire substrates were co-loaded during the growth process. The AFM images in Figure 9 (a) and (b) reveal an RMS roughness of 1.82±0.04 nm for the sample grown on the GaN template, and 5.23±0.31 nm for the sample grown on sapphire. These results indicate that the sample grown on the GaN template has a significantly smoother surface compared to the one grown on sapphire.

In this part, we focus on the SEM and AFM images of MgGeN$_2$ films grown on c-sapphire substrates using both pulsed and continuous growth modes. Figures 7 (h) and (i) present SEM images for the pulsed growth mode, with a germanium concentration of 2000 ppm, and the continuous growth mode, with a germanium concentration of 200 ppm. Both samples were grown on double-sided polished c-sapphire substrates under the same growth temperature and II-IV flow ratio. The SEM images for both growth modes show a similar rough morphology, characterized by faceted surfaces.

Further confirmation of these morphological features comes from the AFM images in Figures 9 (c) and (d), which demonstrate comparable surface roughness between the pulsed and continuous growth modes. However, despite the visual similarities in morphology, EDS data reveal a difference in film thickness. The sample grown using the pulsed growth mode has a thickness of 79±2 nm, whereas the sample grown with the continuous growth mode has a slightly lower thickness of 73±2 nm. This indicates that pulsed growth mode may offer advantages in optimizing growth rates while maintaining surface morphology, potentially making it a more efficient method for synthesizing high-quality MgGeN$_2$ films.

We now compare the XRD results obtained from different growth modes and instruments. Figure 10 shows the XRD 2θ-ω scan profiles for MgGeN$_2$ samples E and K, along with the underlying GaN sample. The GaN sample displays a strong peak at around 2θ = 34.5°, corresponding to the GaN (002) plane. In sample E, distinctive MgGeN$_2$ peaks can be identified, including the (002) peak at around 2θ = 34.64° and the (120) peak at approximately 2θ = 31.56°. Additionally, with the use of a different instrument, the (121) peak, which is close to the (002) peak, is also observed, further confirming the presence of MgGeN$_2$.

## C. Calculated and measured absorption coefficients

A Tauc plot is often used to assess whether a band gap is direct or indirect and to estimate its value. A linear behavior in a Tauc plot corresponds to a $\omega^{1/2}$ dependence above the absorption onset, which in turn is based on a parabolic inter-band difference with fixed curvature. In Figure 11 we show Tauc plots obtained from transmittance measurements on three of our samples. Two straight-line segments are apparent; these extrapolate to about 5.7 and 4.2 eV. A priori it is not clear whether the upper value corresponds to the direct band gap and the lower value to a defect absorption tail, or the lower extrapolated value is the gap. However, the measured magnitude of the absorption in the 4-5 eV range is of order $10^5$ cm$^{-1}$, which would be unusually large for defect absorption. To clarify the situation, we did an ab initio calculation of the band structure in order to generate the calculated Tauc plot, also shown in Figure 11, to compare directly with the experiment.

The band structure calculation was done using the same *QSGW* method as in [10] but with a larger basis set that includes Mg-*2p* and Ge-*3d* semi-core levels treated as local orbitals. The calculation yields a direct band gap of 4.28 eV using the experimental lattice constants [15]. The dielectric function was calculated in the independent particle approximation without inclusion of excitonic effects so that we could use a large k-mesh and the tetrahedron Brillouin zone integration method, which describes the shape of the imaginary part of the dielectric function $\varepsilon_2(\omega)$ more accurately without the need for broadening by an imaginary part of the energy. From $\varepsilon_2(\omega)$ we obtained the real part $\varepsilon_1(\omega)$ by the Kramers-Kronig transformation and then the absorption coefficient is $\alpha(\omega)=\omega \varepsilon_2(\omega)/n(\omega)c$ with n the real part of the index of refraction, with $n+i\kappa = (\varepsilon_1 + i\varepsilon_2)^{1/2}$. Figure 12(a) shows the computed band structure. Figure 12(b) shows the absorption coefficient resulting from averaging $\varepsilon_2(\omega)$ for the polarizations along the crystallographic *a* and *b* axes. This was done for comparison with the experimental absorption spectra, which were measured for unpolarized light propagating along the growth (*c*) axis. Figure 12(c) shows the Brillouin zone with symmetry points labeled.

In Figure 12(a) we see that the conduction band becomes close to linear already slightly above the parabolic minimum. Secondly, three closely spaced valence bands occur at $\Gamma$ with an anisotropic effective mass tensor, which means that within about 0.2 eV these bands cross or have avoided crossings and switch to a different curvature. These details might be better seen if we look directly at $\alpha(\omega)$ in Figure 12(b) instead of at the Tauc plot. Since the top valence band has $a_1$ symmetry and allows transitions to the conduction band minimum (CBM) only for polarization along the c-direction, for the experimental conditions used here the absorption coefficient should be almost negligible for photon energies below about 4.5 eV. At 4.7 eV contributions from the $b_2$ valence band, which crosses the second at about 0.4 eV below the valence band maximum, appear. At about 2 eV above the CBM transitions to the second and third conduction bands along $\Gamma Z$ kick in. The strong absorption from these bands, related to the higher joint density of states, with nearly parallel valence and conduction bands, gives rise to the higher slope part of the Tauc plot. We can thus understand the calculated behavior of the absorption coefficient in detail with respect to the underlying band transitions.

We note that this calculation does not include zero-point-motion or finite temperature electron-phonon coupling corrections, estimated to be of order -0.1 to -0.2 eV [27,28]. The absolute value of the absorption coefficient depends on transition dipole matrix elements which tend to be overestimated in the calculation. Nonetheless in the Tauc plot the theory seems to underestimate the experimental value. On the other hand, in the experimental extraction of the absorption coefficient from the transmittance, uncertainties or systematic errors in the measurement of the thickness of the layer, or scattering arising from roughness effects at the surface and the interface to the transparent sapphire substrate, could lead to smaller transmittance than expected and thus a larger absorption coefficient. RMS surface roughnesses measured by AFM are a few nm, more than an order of magnitude smaller than the shortest optical wavelength, and thus scattering from the surface should be negligible. In addition, the measured transmittance below the band gap is as expected. Roughness at the $MgGeN_2$-sapphire interface could occur through the formation of voids or composition fluctuations associated with the first layers of growth; these cannot be ruled out.

The experimental structures may have some cation disorder, which is expected to lower the gap. Nonetheless, the calculated band gap of 4.28 eV is consistent with the experimental Tauc plots and with a direct gap, and the change in slope of the Tauc plots at higher energies is associated with higher band transitions similar to what was discussed in a recent analysis of the optical properties of $ZnGeN_2$ [29].

## IV. Conclusions

We report here the growth by MOCVD of stoichiometric MgGeN$_2$ films on GaN and c-sapphire substrates using MOCVD. We found that both the growth temperature and the II/IV flow ratio significantly affected the cation composition. Under an idealized model assuming 100% precursor efficiency and negligible re-evaporation, the predicted growth rates based on the Mg and Ge precursor flow rates were 5–6 times and approximately 3.5 times greater, respectively, than the experimental values. Both pulsed and continuous growth modes produced smooth morphologies, appropriate Mg:Ge stoichiometric ratios, and distinct MgGeN$_2$ diffraction peaks in XRD. Optical transmittance measurements of MgGeN$_2$ films grown on sapphire yielded a band gap of $4.28 \pm 0.06$ eV, aligning well with theoretical predictions. These results represent an important step toward the development of Mg-IV-N$_2$ materials—in particular MgGeN$_2$, MgSiN$_2$, and their alloys with binary nitrides—for potential use in deep-ultraviolet optoelectronics and power devices.

**Acknowledgements**
This work was supported by the Army Research Office (Award No. W911NF-24-2-0210).

WL and IT acknowledge support from the U.S. Department of Energy Basic Energy Sciences (DOE-BES) under grant no. DE-SC0008933. Their calculations of the band structure and optical properties made use of the High Performance Computing Resource in the Core Facility for Advanced Research Computing at Case Western Reserve University.

Use of the x-ray diffractometer at Oberlin College was supported through NSF DMR-MRI 0922588.

NH acknowledges financial support from the Case School of Engineering for her contributions to the XRD analysis.

# Author Declarations

## Conflict of Interest

The authors have no conflicts of interest to disclose.

## Data Availability

The data that support the findings of this study are available from the corresponding author upon reasonable request.

**Table Captions**

**Table 1.** Growth parameters. For samples B and D, t_on included a 24 s ramp-up and 24 s ramp-down of the germane flow rate, followed by a 12 s off time. For samples C, E, and F, t_on includes 25 s for both ramp-up and ramp-down of the germane flow rate, with a subsequent 10 s off time. The $GeH_4$ flow rates were set at 20 sccm (1.8 micromoles/min) for samples B and D, 27 sccm (2.4 micromoles/min) for samples C and E, and 37 sccm (3.3 micromoles/min) for sample F. For samples A-D, the $NH_3$ flow rate was 3000 sccm. For the rest of the samples the $NH_3$ flow rate was 1500 sccm. For samples A-E the Mg precursor bubbler temperature was held at 30 °C. For samples F-K the bubbler temperature was 40 °C. The chamber pressure was 500 Torr for all samples.

**Table 2.** Parameters used in the calculation of growth rates based on Mg or Ge precursor flow rates, listed along with the calculated and measured growth rates.

**Table 1**

| Sample | Substrate | Pyro T (°C) | Cp$_2$Mg (μmol/min) | MeCp$_2$Mg (μmol/min) | t (min) | Growth method | R$_{II/IV}$ | Ga% | Thickness (nm) | Mg/Ge |
|---|---|---|---|---|---|---|---|---|---|---|
| A(#016) | GaN | 700 | 1.5 | - | 120 | Continuous | 0.1 | 19.7 ± 2.0 | - | 0.55 ± 0.10 |
| B(#021) | GaN | 815 | - | 7.1 | 120 | Pulsed | 10 | 9.2 ± 1.0 | - | 1.12 ± 0.18 |
| C(#022) | GaN | 745 | - | 7.1 | 120 | Pulsed | 7 | 10.3 ± 0.6 | 96 ± 3 | 1.13 ± 0.10 |
| D(#023) | GaN | 815 | - | 7.1 | 120 | Pulsed | 10 | 11.8 ± 0.7 | 90 ± 3 | 1.00 ± 0.09 |
| E(#028) | GaN | 745 | - | 7.1 | 120 | Pulsed | 7 | 2.6 ± 0.2 | 130 ± 5 | 0.99 ± 0.08 |
| F(#029) | GaN | 745 | - | 9.7 | 60 | Pulsed | 7 | 7.3 ± 0.4 | 105 ± 4 | 1.00 ± 0.09 |
| G(#038) | Sapphire | 745 | - | 9.7 | 60 | Pulsed | 13 | - | 79 ± 2 | 0.98 ± 0.08 |
| H(#039) | Sapphire | 745 | - | 9.7 | 60 | Continuous | 13 | - | 73 ± 2 | 0.99 ± 0.08 |
| I(#041) | GaN | 745 | 3.6 | - | 60 | Continuous | 9 | 28.7 ± 1.4 | 36 ± 1 | 1.05 ± 0.09 |
| J(#042) | GaN | 745 | 3.6 | - | 120 | Continuous | 9 | 14.2 ± 0.7 | 68 ± 2 | 1.04 ± 0.09 |
| K(#044) | Co-load | 745 | 3.6 | - | 120 | Continuous | 9.2 | 12.1 ± 0.6 | 86 ± 2 | 1.07 ± 0.09 |

**Table 2.**

| Sample | T (°C) | $P^*_{pr\_Ge}$ (Pa) | $v_{N_2}$ (10$^{-6}$m$^2$/s) | $D_{AB\_Ge}$ (10$^{-5}$m$^2$/s) | Growth rate_Ge (nm/hour) | Growth rate_Mg (nm/hour) | Experimental Growth rate (nm/hour) |
|---|---|---|---|---|---|---|---|
| C | 745 | 0.27 | 191.5 | 28.9 | 181.4±13.6 | 241.0±78.3 | 48±1 |
| D | 815 | 0.20 | 213.7 | 31.9 | 126.3±9.5 | 236.0±76.7 | 45±1 |
| E | 745 | 0.38 | 191.5 | 28.9 | 250.0±18.8 | 332.3±107.9 | 65±2 |
| F | 745 | 0.52 | 191.5 | 28.9 | 342.2±25.7 | 455.6±147.9 | 105±5 |
| G | 745 | 0.28 | 191.5 | 28.9 | 184.6±13.9 | 269.2±87.4 | 79±2 |
| H | 745 | 0.28 | 191.5 | 28.9 | 183.3±13.8 | 267.8±86.9 | 73±2 |
| J | 745 | 0.15 | 191.5 | 28.9 | 99.2±7.4 | 204.9±65.7 | 34±1 |
| K | 745 | 0.15 | 191.5 | 28.9 | 97.0±7.3 | 205.0±65.7 | 43±1 |

# Figure Captions

**Figure 1.** (a) Calculated band gaps versus the wurtzite lattice parameter $a_w$ for MgGeN$_2$, MgSiN$_2$, GaN and AlN. (b) Calculated band gaps versus the orthorhombic lattice parameters $a_0$ and $b_0$ for MgGeN$_2$, MgSiN$_2$, GaN and AlN. The open symbols represent the data for the indirect band gap of MgSiN$_2$. The solid symbols represent direct band gaps of the materials. For consistency of comparison, the lattice parameters for all four materials were taken from the R2SCAN calculations listed on the Materials Project website. The wurtzite lattice parameter $a_w$ for GaN and AlN was translated to the orthorhombic lattice parameters $a_o$ and $b_o$ using the relationships $a_0 = 2a_w$ and $b_0 = \sqrt{3}a_w$.

**Figure 2.** Schematic diagram of the steps in each cycle in the pulsed-mode growth.

**Figure 3.** The Mg/Ge ratio measured by EDS versus the ratio $R_{II/IV}$ of the flow rates of the Mg precursor to germane for samples grown at three different growth temperatures (700 °C, 745 °C, and 815 °C). All other growth parameters were the same. The higher $R_{II/}$ ratio required to achieve stoichiometry for the growth at 800 °C is likely due to the higher evaporation rate of Mg.

**Figure 4.** Cross-sectional STEM images of sample E showing (a)-(c) the uniform distribution of the different elements, (d) the line scan for (e) the measurement of the atomic ratios across the interface between the MgGeN$_2$ film and the GaN layer.

**Figure 5.** (a) The measured EDS spectrum of sample E compared with the simulated spectrum. The thickness of the MgGeN$_2$ layer (130 nm) was chosen to best fit the ratio of the Ga to the Mg and Ge peaks. (b) The measured EDS spectra for sample E and sample F at an accelerating voltage of 5keV and beam current of 3.2 nA.

**Figure 6.** Calculated and experimental growth rates for samples C, D, E, F, G, H, J and K.

**Figure 7.** Plan-view FESEM images of MgGeN$_2$ films grown under various conditions. Panels (a–c) show samples grown using MeCp$_2$Mg as the Mg precursor: (a) sample D grown at 800 °C, (b) Sample E grown at 720 °C with a 30 °C Mg bubbler temperature, and (c) sample F grown at 720 °C with a 40 °C Mg bubbler temperature. Panels (d–g) show samples grown using Cp$_2$Mg: (d) sample I, (e) sample J, (f) sample K grown on GaN, and (g) sample K grown on sapphire. Finally, (h) sample G was grown in pulsed mode using MeCp$_2$Mg, and (i) sample H was grown in continuous mode using MeCp$_2$Mg.

**Figure 8.** Plan-view FESEM images and EDS results for different spots on t (a) sample I and (b) sample J.

**Figure 9.** Plan-view AFM images (a) sample K on GaN, (b) sample K on sapphire, (c) sample G (pulsed growth mode) and (d) sample H (continuous growth mode).

**Figure 10.** XRD 2θ–ω scan profiles for sample E, sample K, and an unintentionally doped -GaN substrate. Spectrum (a) represents the GaN XRD scan from Oberlin, while spectrum (b) shows the XRD scan of sample E from Oberlin. Spectra (c), (d), and (e) correspond to GaN, sample E, and sample K XRD scans, respectively, from Case Western Reserve University.

**Figure 11.** Tauc plots of three stoichiometric MgGeN$_2$ samples (G, H, K), along with the calculated Tauc plot. Sample G was grown using the pulsed-flow method, while samples

H and K employed continuous flow. For clarity, the calculated curve has been multiplied by 2.5 and 10, both indicated at the labeled points.

**Figure 12.** (a) Computed conduction and valence band structure, (b) calculated absorption coefficient as a function of photon energy, and (c) schematic of the Brillouin zone with labeled high-symmetry points.

**Figure 1.**

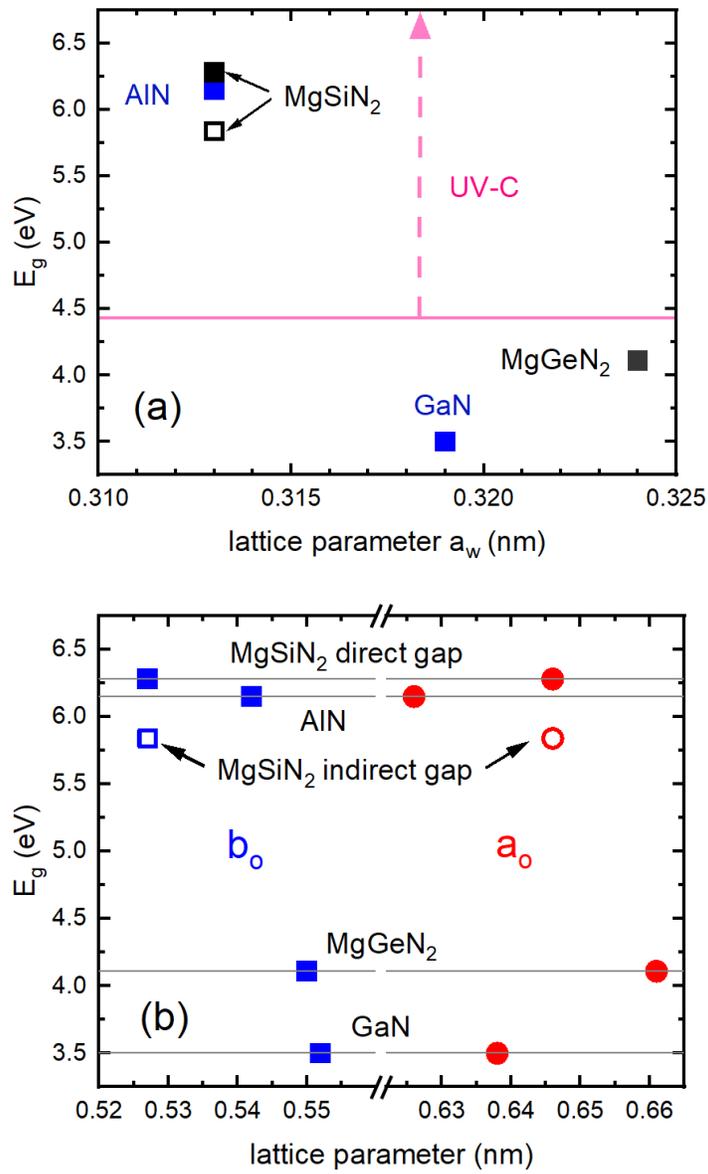

**Figure 2.**

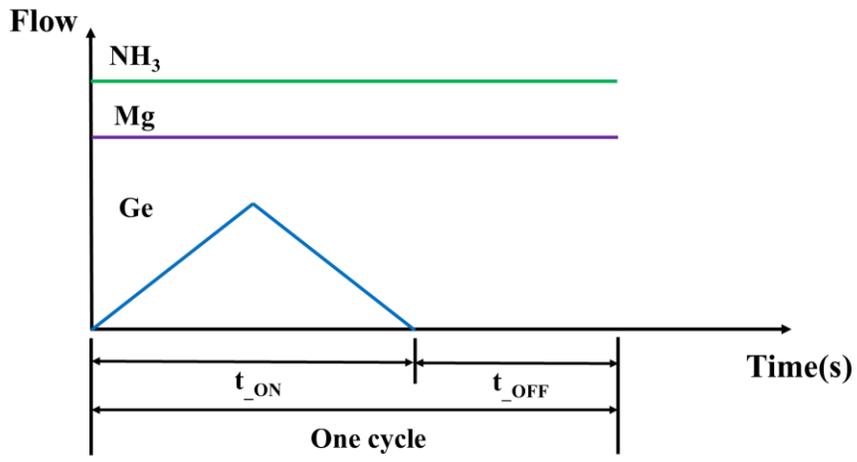

**Figure 3.**

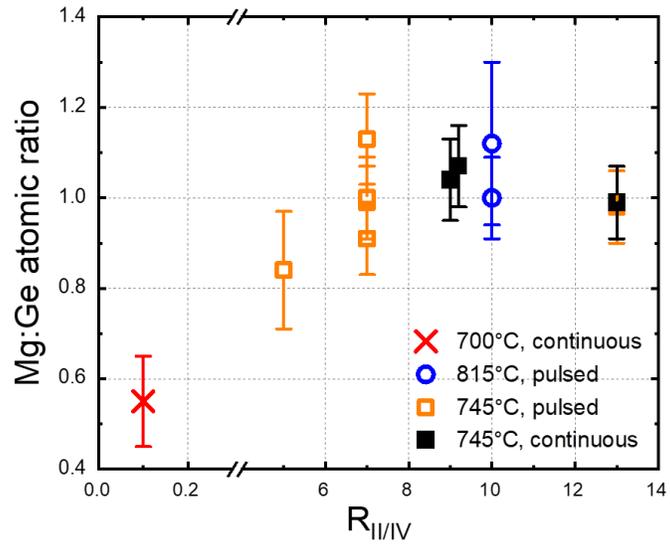

**Figure 4.**

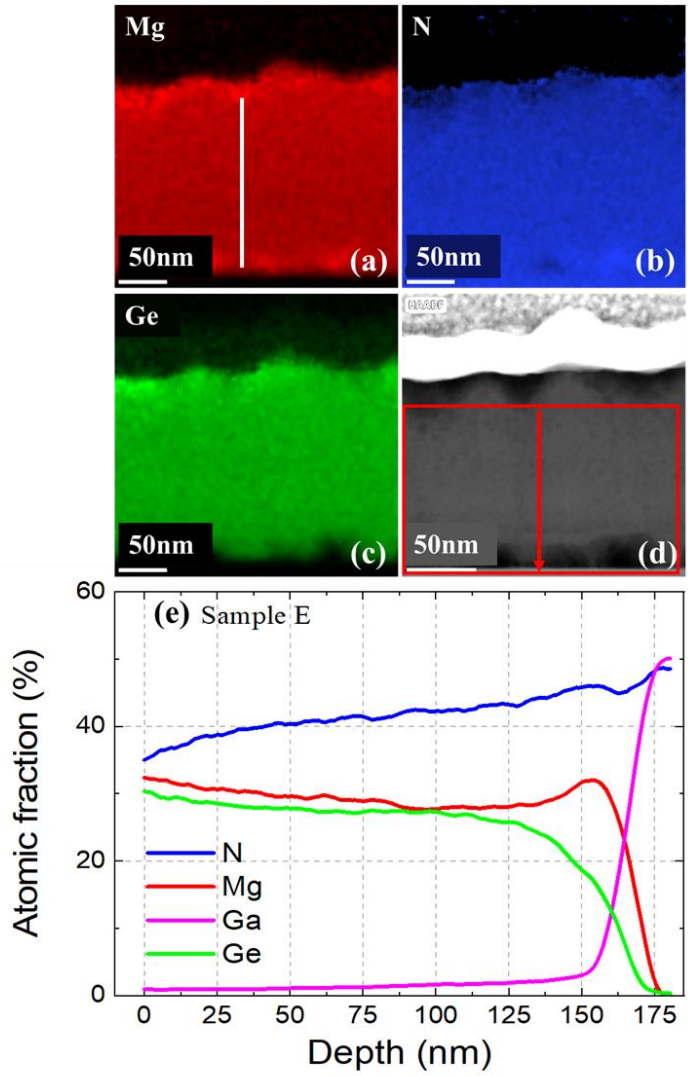

**Figure 5.**

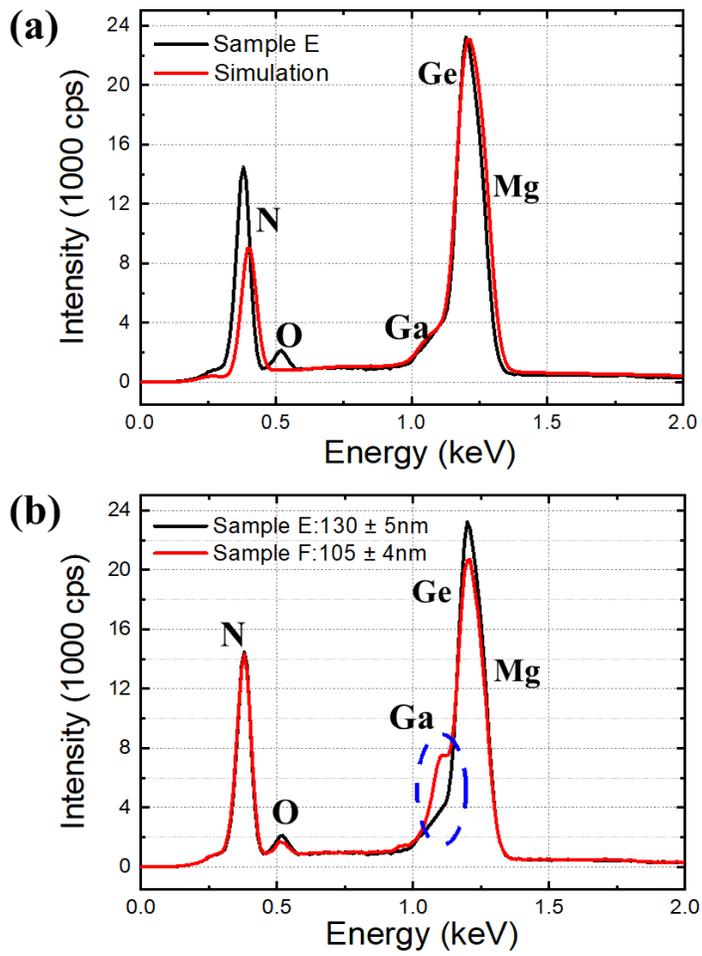

**Figure 6.**

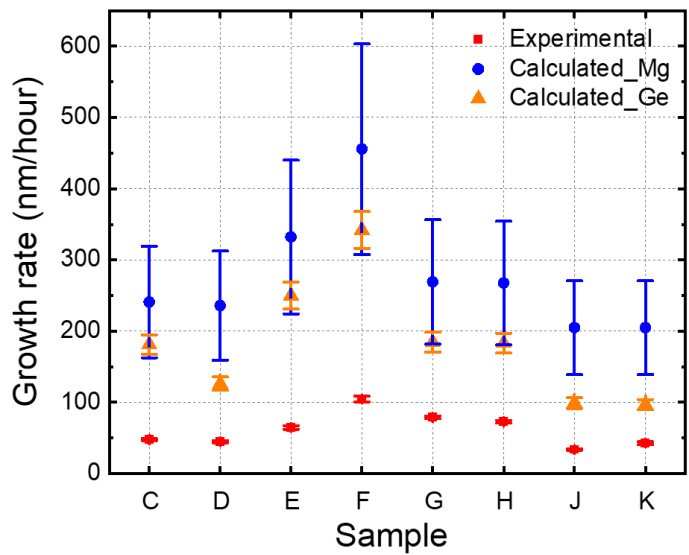

**Figure 7.**

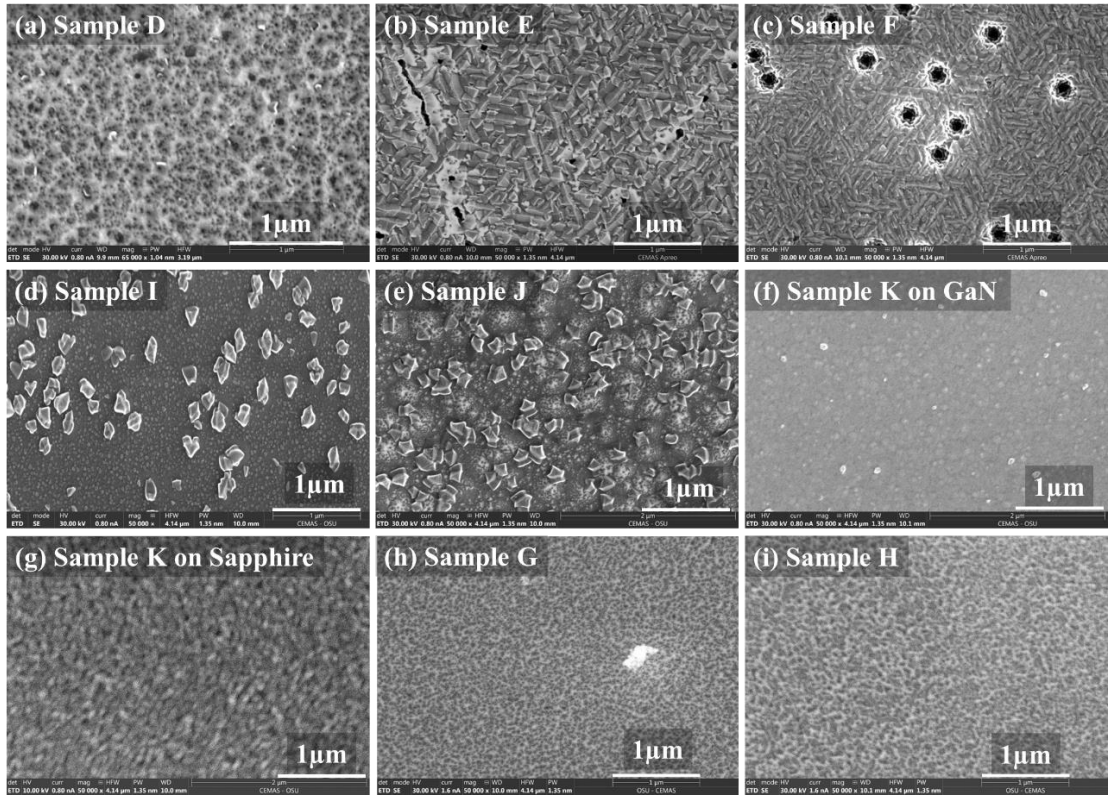

**Figure 8.**

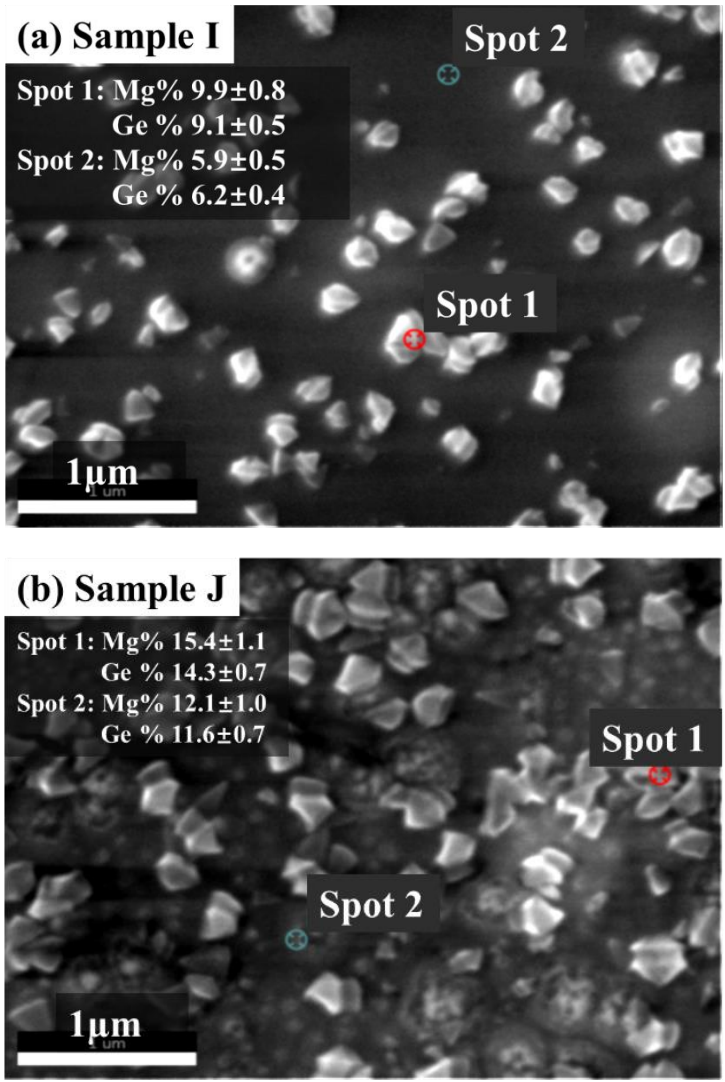

**Figure 9.**

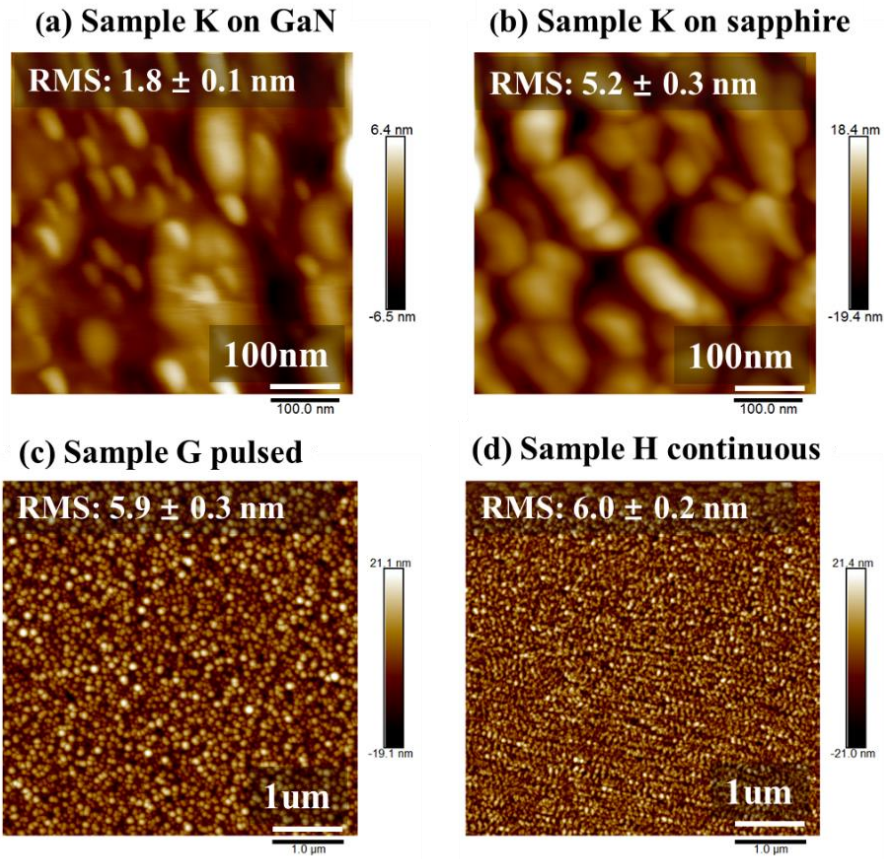

**Figure 10.**

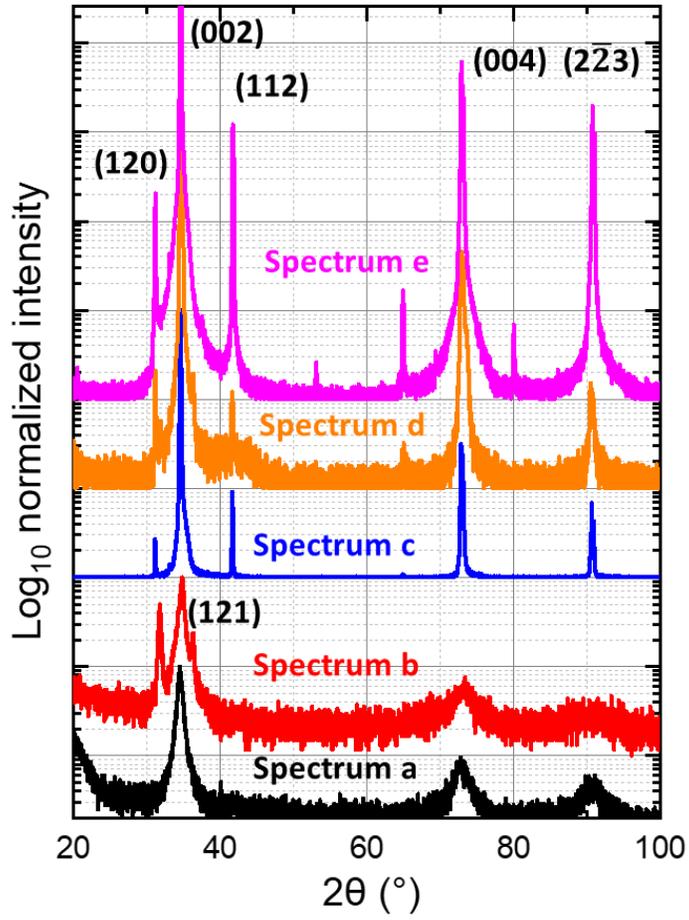

**Figure 11.**

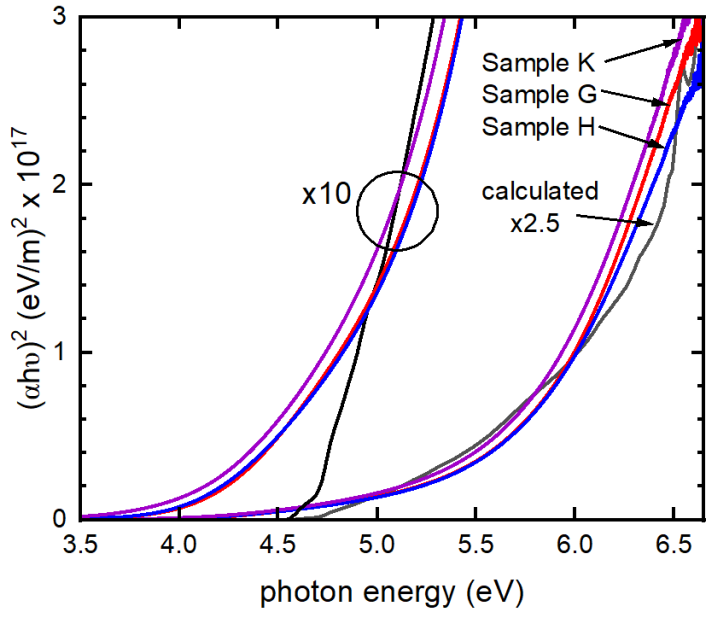

**Figure 12.**

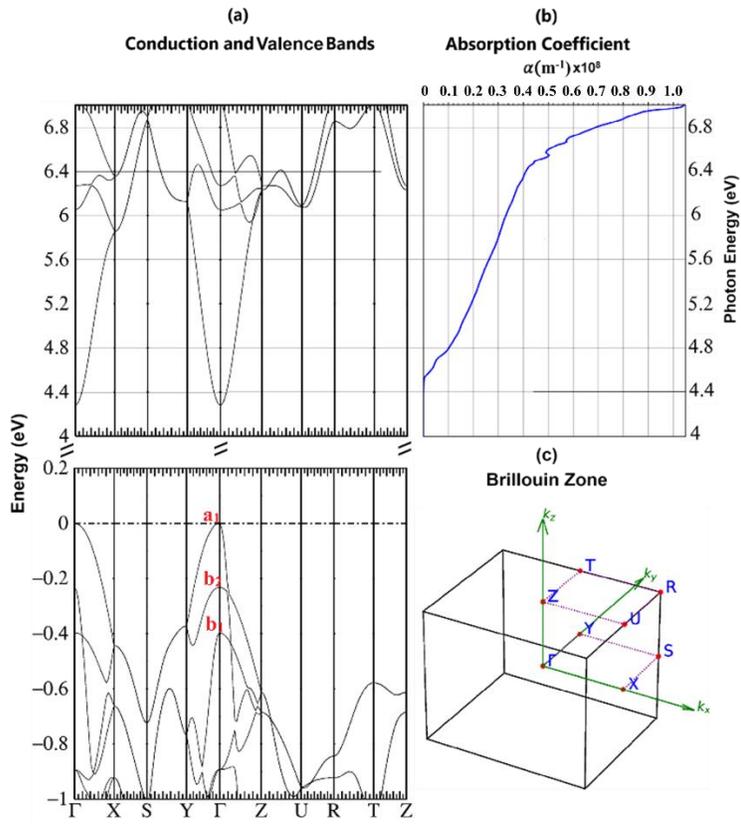